\documentclass[aps,prl,twocolumn,superscriptaddress,preprintnumbers,floatfix,nofootinbib]{revtex4}
\usepackage{graphicx}
\usepackage{epsfig}

\begin{document}

\newcommand{\beq}{\begin{eqnarray}}
\newcommand{\eeq}{\end{eqnarray}}
\newcommand\del{\partial}
\newcommand{\nn}{\nonumber } 
\newcommand{\tK}{Z_{n=1}}
\newcommand{\re}{\mathrm{Re}}
\newcommand{\im}{\mathrm{Im}}
\newcommand{\Str}{\rm Str}
\newcommand{\bmat}{\left ( \begin{array}{cc} 
}
\newcommand{\emat}{\end{array} \right )}
\newcommand{\hm}{\hat{m}}
\newcommand{\hz}{\hat{z}}
\newcommand{\hx}{\hat{x}}
\newcommand{\ha}{\hat{a}}
 
\title{Constraints on New Physics from Baryogenesis and Large Hadron Collider Data}

\author{Poul~H.~Damgaard}
\affiliation{Niels Bohr International Academy and Discovery Center, 
Niels Bohr Institute, University of Copenhagen, Blegdamsvej 17, DK-2100 Copenhagen, Denmark}

\author{Donal~O'Connell}
\affiliation{Niels Bohr International Academy and Discovery Center, 
Niels Bohr Institute, University of Copenhagen, Blegdamsvej 17, DK-2100 Copenhagen, Denmark}

\author{Troels C. Petersen}
\affiliation{Niels Bohr International Academy and Discovery Center, 
Niels Bohr Institute, University of Copenhagen, Blegdamsvej 17, DK-2100 Copenhagen, Denmark}

\author{Anders Tranberg}
\affiliation{Niels Bohr International Academy and Discovery Center, 
Niels Bohr Institute, University of Copenhagen, Blegdamsvej 17, DK-2100 Copenhagen, Denmark}
\affiliation{Faculty of Science and Technology, University of Stavanger, N-4036 Stavanger, Norway}

\date   {\today}

\begin{abstract}
We demonstrate the power of constraining theories of new physics by insisting that they lead to electroweak baryogenesis, while agreeing with current data from the Large Hadron Collider.
The general approach is illustrated with
a singlet scalar extension of the Standard Model. Stringent bounds can already be obtained, 
which reduce the viable parameter space to a small island.
\end {abstract}

\maketitle

\noindent {\it Introduction.} 
The possibility of naturally explaining the observed baryon asymmetry of the universe
within the Standard Model of particle physics was thwarted by a combination
of analytical efforts and experimental data from LEP~\cite{Barate:2003sz}. Otherwise, 
the ingredients seemed
to line up nicely: CP-violation, sphaleron transitions, and the possibility of a period
in the early universe where a first order electroweak phase transition could ensure the
required 
departure from equilibrium 
that is one of Sakharov's original conditions \cite{Kuzmin} (for a review, 
see \cite{Rubakov}). The last ingredient
was crucial, and its availability seemed to be indicated by naive tree-level estimates in
a large span of Higgs masses. Nevertheless, this attractive explanation of baryogenesis
was foiled when it was realised that a sufficiently strong first order phase transition requires a lighter Higgs particle than tree-level estimates had suggested~\cite{Kari}.
This lower bound
was quickly reached by the LEP experiment. Therefore, the
possibility of explaining the observed baryon asymmetry in the Standard Model alone was ruled out. This
is an astonishingly powerful conclusion which immediately requires
physics beyond the Standard Model.

In this paper, we describe a general approach for constraining new physics theories, using data
from the Large Hadron Collider (LHC) and imposing the additional constraint that electroweak baryogenesis must be viable~\cite{Menon:2009mz}.
This approach is very attractive, for the following reason. Models of new physics can typically evade LHC bounds by reducing the coupling between new degrees of freedom and the Standard Model particles. On the other hand, electroweak baryogenesis only occurs when the new physics interacts with the Higgs sector strongly enough to change the nature of the phase transition. Thus, we corner the model. Indeed, as we shall see below, our general approach allows us to reduce the allowed parameter space of an illustrative model to a small island, which will be further reduced when new LHC data becomes available. 

Although we are forced to consider new physics, we restrict our attention to models for which
the mechanism for generating baryon number asymmetry relies solely on 
electroweak physics, {\em except} for new physics that will ensure a strong first
order transition and additional CP-violation \cite{shapCP}.
This seems a natural minimalistic stance, and one that preserves all the
positive features of the original idea of (Standard Model) electroweak baryogenesis. The only new ingredients are one or more fields that couple to the electroweak
sector.
Thus, the asymmetry is tied to the chiral anomaly in the electroweak sector, as in the Standard Model.
Completely different mechanisms for generating
baryon asymmetry from physics beyond the Standard Model
(such as leptogenesis, see, $e.g.$, Ref.~\cite{Buchmuller:2005eh} for a review) are possible. We do not consider such models here.

\noindent {\it A singlet scalar extension.}
The simplest theory of the kind
we wish to analyse is the singlet extension of the Standard Model 
(see for instance 
Ref.~\cite{Donal}). 
In this model, only one new field, 
a gauge singlet real scalar $S(x)$, is introduced. We consider all possible renomalisable couplings between the $S$ and the Standard Model fields. The $S$ can only couple through the Higgs portal; therefore,
the model is completely specified by the scalar sector, which is
\beq
{\cal L} = (D_{\mu}H(x))^{\dagger}D^{\mu}H(x) + \frac{1}{2}(\partial_{\mu}S(x))^2
- V(H,S) ,
\eeq
where $D_{\mu}$ is the usual covariant derivative, $H$ is the Standard Model Higgs doublet, and the potential $V(H,S)$ is
\begin{eqnarray}
&&V(H,S)={m^2 \over 2} H^{\dagger}H +{\lambda \over 4} (H^{\dagger}H)^2 +{\delta_1 \over 2}H^{\dagger}HS  \cr
&&\!\!+{\delta_2 \over 2}H^{\dagger}HS^2+{\delta_1 m^2 \over 2 \lambda}S+
{\kappa_2 \over 2} S^2 +{\kappa_3 \over 3}S^3+{\kappa_4 \over 4}S^4. 
\end{eqnarray}
The couplings $\delta_1,\delta_2,\kappa_2,\kappa_3,\kappa_4$ are
real, but are otherwise not constrained beyond the requirement of a bounded potential. Without loss of generality, we have chosen parameters so that the vev of $S$ vanishes~\cite{Donal}. There
are no tree-level couplings between the scalar field $S(x)$ and the other fields of the Standard
Model, and in particular there are no new sources of CP violation. Such sources are required for the baryon asymmetry to be large enough, and can for instance be introduced through higher dimension operators as in Ref. \cite{Cline}. For our purpose here, this will not be relevant.
What will be relevant is that this simple extension of the Standard Model can lead to 
a radically different pattern of first order finite temperature phase transitions; this fact 
was noted early in \cite{Hall}.

Although the model introduces several new fundamental parameters, the physics is simple, both from the point of view of LHC phenomenology and baryogenesis. The key observation is that the $S$ interacts with the Standard Model only via the Higgs. Indeed, when $\delta_1 \neq 0$, the $S$ and the Higgs scalar mix. 
The mixing angle and the mass of the new scalar state are the key parameters controlling the LHC signals of the model. The mass matrix can be read off from the quadratic terms of the scalar potential after expanding about the Higgs vev:
\begin{equation}
V_{ \rm mass}={1 \over 2}\left(\mu_h^2 h^2 +\mu_S^2 S^2 +\mu_{hS}^2 h S \right),
\end{equation}
where
\begin{equation}
\label{param}
\mu_h^2=\frac{m^2}{2}+\frac{3\lambda v^2}{4},\quad
\mu_S^2= \kappa_2+\frac{\delta_2 v^2}{2} , \quad
\mu_{hS}^2=\delta_1 v.
\end{equation}
To identify the physical scalar degrees of freedom we must diagonalize the mass matrix. There are two eigenstates, one of which must correspond to the observed $\sim 125$ GeV state. We will denote this state by $h_{125}$. We will primarily be interested in the case where the other state, $\varsigma$, is the heavier of the two neutral scalars. 
These fields are
simply linear combinations of the Standard Model Higgs scalar field $h$ and the new field $S$. 
Explicitly, the field $\varsigma$ has mass
\begin{equation}
m_\varsigma^2 = \frac{\mu_h^2 + \mu_S^2}{2} - \frac{\mu_h^2 - \mu_S^2}{2} \sqrt{1 + x^2} .
\end{equation}
In terms of the original fields in the Lagrangian, $\varsigma$ is given by
$
\varsigma = {\rm cos}\,\theta ~S- {\rm sin}\,\theta ~h,
$
where~\cite{Krasnikov}
\begin{equation}
\tan \theta={ x \over 1+\sqrt{1+x^2}},~~~x={\mu_{hS}^2 \over \mu_h^2 -\mu_S^2}.
\end{equation}
Notice that the parameters $\delta_1$ and $\delta_2$ play an important role in determining the scalar masses and mixings. Since $\delta_1$ and $\delta_2$ are the only parameters coupling the $S$ to the Higgs, it is easy to anticipate the LHC and baryogenesis constraints may complement each other in this model.

In view of the consistency of the $\sim125$ GeV
state observed at the LHC \cite{pl1}
with the Standard Model Higgs, it is natural to assume that the $h_{125}$ is mostly $h$. 

When the $\varsigma$ is heavier than twice the $h_{125}$ mass, the decay $\varsigma \to h_{125} h_{125}$ is kinematically allowed. 
All other partial widths of the $\varsigma$ are equal to the partial width of a Standard Model Higgs, with mass $m_\varsigma$, times a mixing factor $\sin^2 \theta$. It is convenient to introduce the ratio
\begin{equation}
f(m_\varsigma) = \frac{\sin^2 \theta\, \Gamma^\mathrm{SM}_\mathrm{total}(m_\varsigma)}{\Gamma_\mathrm{total}(m_\varsigma)},
\end{equation}
where $\Gamma^\mathrm{SM}_\mathrm{total}(m_\varsigma)$ is the total width of a Higgs scalar with mass $m_\varsigma$ in the Standard Model, while $\Gamma_\mathrm{total}(m_\varsigma)$ is the width of the $\varsigma$ in the present model. Then, the branching ratios of the $\varsigma$ to final states $\psi$ other than $h_{125} h_{125}$ are simply given by
\begin{equation}
\mathrm{BR}(\varsigma \rightarrow \psi) = f(m_\varsigma) \mathrm{BR}^\mathrm{SM}(h \rightarrow \psi).
\end{equation}
Meanwhile, the branching ratio of the heavier scalar into a pair of the lighter scalars is
\begin{equation}
\mathrm{BR}(\varsigma \rightarrow h_{125} h_{125}) =1 - f .
\end{equation}
A short calculation shows that the ratio $f$ is given, at tree level, by
\begin{equation}
f = \left( 1 + \frac{\cot^2 \theta }{128 \pi} \sqrt{1 - \frac{4 m_{125}^2}{m_\varsigma^2} } \sin^2 2 \theta \frac{(m_{125}^2 -m_\varsigma^2)^2}{ m_\varsigma v^2 \Gamma^\mathrm{SM}_\mathrm{total}(m_\varsigma)}  \right)^{-1},
\end{equation}
when $m_\varsigma \geq 2 m_{125}$, and 
$f=1$
otherwise.
The decays
$\varsigma \to W W$ and $\varsigma \to Z Z$ dominate even above the two-$h_{125}$
threshold, which gives rise to only a small change in the partial widths. 


\noindent {\it Searching for first order phase transitions.}
Our first goal is to determine the region of parameter space
that is compatible with a strongly first order phase transition.

We compute the one-loop finite temperature effective potential of the Standard Model + singlet, in outline following~\cite{Espinosa, Kimmo}. Schematically, the potential can be written
\begin{equation}
V_{\rm eff}(T)= V_{\rm c}+V_{\rm CW}+V_{\rm ct}+V_{\rm T},
\end{equation}
in terms of the tree-level contribution (c); the ($T=0$) Coleman-Weinberg potential (CW); counter terms fixed so that the renormalised potential reproduces the tree-level minima, masses and couplings at $T=0$ (ct); and the finite temperature contribution (T) (see for instance \cite{Espinosa}). We include all quark species. 

For each value of the temperature we determine the minimum of the potential in $h, S$-space. This has the advantage that the 
minimum of the effective potential is well-defined and real-valued throughout, whereas it may acquire an imaginary part away from the minimum. A first order transition corresponds to a discontinuous jump in the location of the Higgs field minimum $v(T)$ as temperature is increased. The critical temperature $T_c$ is the temperature where the jump occurs. For a second order transition, $v(T)$ goes smoothly to zero with increasing temperature. 
At the level of our approximation, we see no cross-over transitions. 

\begin{figure}[t!]
  \unitlength1.0cm
 \includegraphics[width=9.2cm]{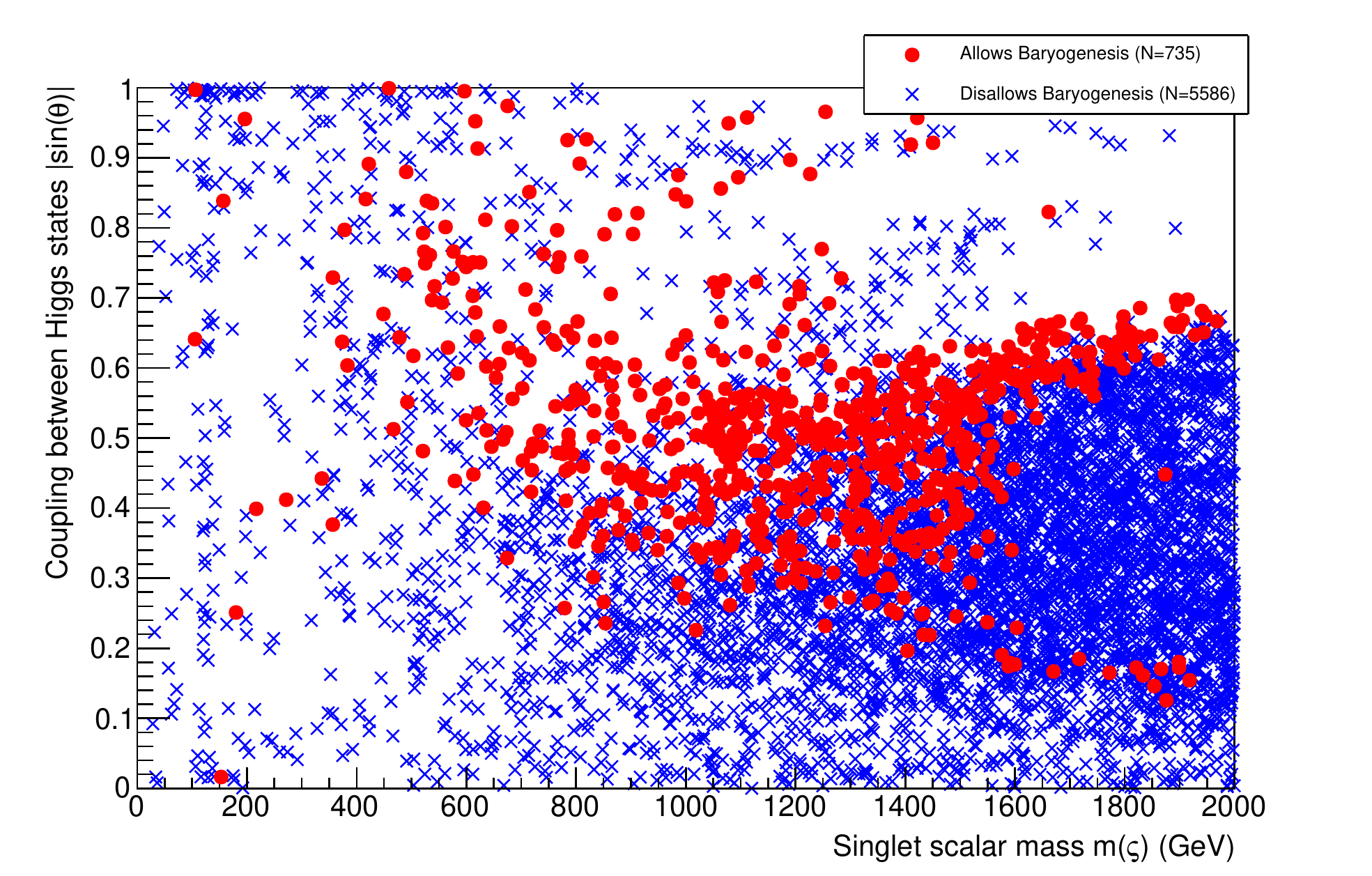}
  \caption{ \label{fig:scatter} A scan of parameter space leading to strong enough first order
(red bullets) and too weak or second-order phase transitions (blue crosses).}
  \vspace{-4mm}
\end{figure}

We follow a more direct numerical approach than \cite{Espinosa} where we make a very large numerical
scan over parameter space, directly identifying all points in this parameter space that
correspond to first order phase transitions of sufficient
strength. Instead of scanning in $\delta_{1,2}$, $\kappa_{2,3,4}$, we use 
the parametrisation in terms of the mass eigenvalues $m_{125} \approx 125$ GeV, and $m_\varsigma$, the mixing angle $\theta$ and four other parameters described in \cite{Espinosa}. We scan uniformly in these parameters, from $-1$ to $1$ in the case of the dimensionless parameters, and from $0$ to $2$ TeV for massive parameters. The dimensionless parameters are (generalised) couplings, and we wish to stay in the perturbative regime. We project the parameter scan onto the plane of $m_\varsigma$, $\left|\sin \theta \right|$, as shown in Fig. \ref{fig:scatter}. Although at the outset the Monte Carlo sampling is uniform, a large number of parameter sets are discarded based on stability of the potential, and also the finite range in which we can track any expectation value the $S$ may develop. We took this range to be $2$ TeV.
These discarded sets are not shown in the figure. To compare to the parameters $\delta_{1,2}, \kappa_{2,3,4}$ in the Lagrangian, this expectation value must of course be removed by redefining the origin of $S$.

We adopt the convention that a phase transition is strongly first order when
\begin{equation}
v(T_c)/T_c ~>~ 0.7.
\end{equation}
Models which satisfy this criterion are shown as red symbols in Fig. \ref{fig:scatter}, while weaker transitions are shown as blue symbols.
Qualitative features of the figure can be understood on intuitive grounds: when the additional scalar has a very large mass, its effect on the potential again fades as dictated by its decoupling.
When the mixing angle is small, we generically cannot find first order transitions except if the coupling gets rather large.

Identifying the precise region in which a sufficiently strong first order phase transition occurs is very challenging and our one-loop approximation introduces some systematic error. 
For example, it is well-known that non-perturbative effects can affect the strength of the phase transition \cite{Kari}. 
So it is reassuring that experience shows that the region of first order phase transitions
{\em shrinks} as approximations are improved, making our criterion that of a conservative estimate.
Another source of systematic error relates to
the exact criterion imposed on $v(T_c)/T_c$ \cite{Patel:2011th}. Restricting to transitions with $v(T_c)/T_c > 1.0$ removes around 7 percent of the transitions. Finally, we examined the theory in the Standard Model-like region near $\theta=0$. As expected, we find second order transitions in this region, with critical temperature within 15 percent of the results obtained in a fully non-perturbative calculation in the Standard Model (albeit at a Higgs mass of 120 GeV) \cite{Kari}. In \cite{Cline}, strongly first order transitions were found in a simplified potential, in a region which in our parametrisation corresponds to $\theta=0$, $m_\varsigma \sim 100$GeV. Indeed, at larger couplings ($\lambda \sim 3$) we generally reproduce the result shown in Fig. \ref{fig:scatter}, except for an additional small island of first order phase transitions at very small mixing angles.

\noindent {\it Bounds from new LHC data.}
Having determined the region of first order phase transitions, we now turn to
present experimental constraints from the LHC.
The main constraint comes from the total Higgs production cross section, which has been
measured to be $0.99 \pm 0.11$ (weighted combination of ATLAS and CMS data~\cite{LHCcouplings}, consistent with general Higgs fits~\cite{invisible}).
This information leads to a constraint on the mixing angle between the $S$ and the Higgs scalar, which is independent of the mass of the heavier scalar state.

\begin{figure}[t!]
  \unitlength1.0cm
\includegraphics[width=9.2cm]{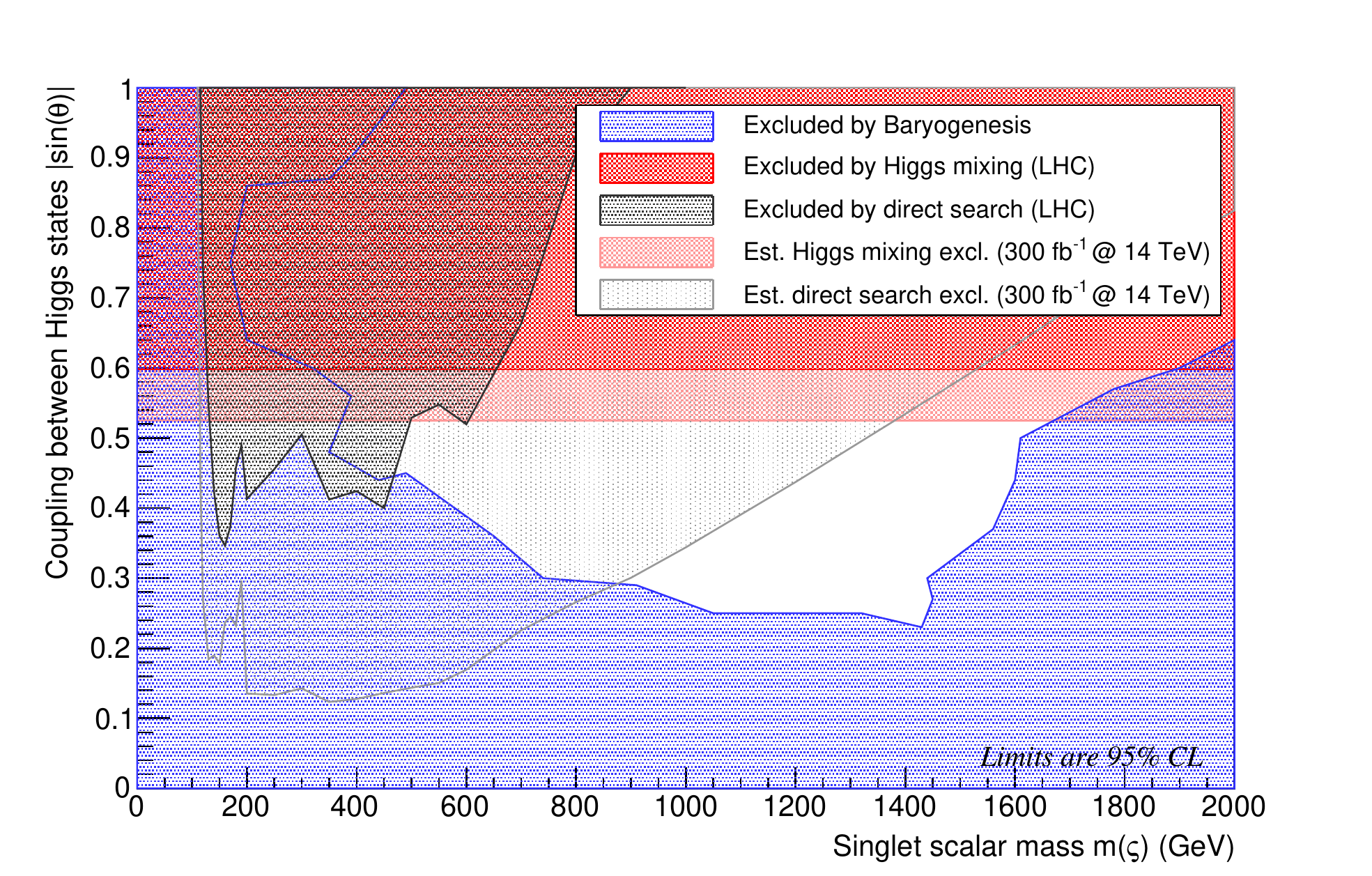}
  \caption{ \label{fig:LHCdata} Experimental bounds from current data of the latest LHC runs, and the exclusion region inferred from Fig.\ \ref{fig:scatter}.}
  \vspace{-4mm}
\end{figure}

We present our results for the allowed region in the mass/mixing angle plane in Fig.~\ref{fig:LHCdata}.
The figure shows that the bound on the mixing angle
dominates the experimental constraints, globally. 
As expected,
it will be hard to rule out small mixing angles from LHC data. However, Fig. \ref{fig:LHCdata}
also illustrates that small mixing angles are irrelevant
in this context, since they do not lead to the required first order phase transition.
Only a narrow region remains. 

There is an additional constraint from direct searches for additional scalars, as a heavy $\varsigma$
would be observable through its decay products.
The direct search constraint is based on recent LHC data~\cite{LHCsearches}, and as can be seen in Fig. \ref{fig:LHCdata}, it is not currently particularly powerful. 

Future running at the LHC will lead to improved constraints on the present model. We have estimated constraints from 300 fb$^{-1}$ of data at 14 TeV, using numbers taken from the recent
feasibility study~\cite{ATLASestimates}. The anticipated future bounds are also shown in Fig. ~\ref{fig:LHCdata}. According to these estimates, which are deliberately conservative, one sees that the full parameter space of the model will not be covered, but the expected new bounds are nevertheless impressive.

\noindent {\it Conclusions.} 
We have demonstrated the power of combining LHC data with constraints from baryogenesis.
We believe the criterion of viable electroweak baryogenesis is on a sufficiently strong
foundation that it may become a standard experimental constraint on proposals for new physics.

Our general approach has been illustrated in detail for what is probably the minimal extension
of the Standard Model in this context: the addition of a single scalar field transforming
trivially under all gauge groups. We have seen that the LHC constraints, in combination with baryogenesis, reduce the viable parameter space of this model to a small region, assuming that coupling constants are small.  This is intuitively sensible; a strongly first order transition requires substantial mixing between the Higgs scalar and the singlet mode; this mixing is highly constrained since the 125 GeV scalar behaves much as expected for a minimal Higgs scalar. In an extended parameter space allowing for larger couplings, specific mechanisms may lead to a strong first order phase transitions; an example is the $Z_2$-symmetric case of \cite{Cline}. As this work was being completed, a related analysis \cite{newwork} appeared for the $Z_2$ symmetric case, emphasising the possibility of generating dark matter by introducing an additional fermion coupling only to $S$. Work is presently underway to extend our analysis to a broader parameter space and 
to the two Higgs doublet model.

Although a first order scenario is only one of the ingredients required for electroweak
baryogenesis, and in particular more sources of CP-violation may be needed \cite{Shu},
we have demonstrated how powerful the first order criterion can be in eliminating the
potential parameter space of new physics theories. When data-taking at the LHC begins again at
almost twice the center-of-mass energy, the present analysis will 
further reduce the parameter space of the singlet scalar model, and will powerfully constrain other theories of physics beyond the Standard Model.

\noindent
{\bf Acknowledgments:}
This work was supported in part by a Lundbeck Junior Group Leader grant (TP) and a
Villum Young Investigator Grant (AT).

\end{document}